\documentclass[a4paper,12pt]{article}
\usepackage[latin1]{inputenc}
\usepackage{amssymb}
\title{{\bf An $\om$-Power of a  Finitary  Language Which is a Borel Set of Infinite Rank} }
\author{Olivier Finkel\\{\it Equipe de Logique Math\'ematique }  
 \\ U.F.R. de Math\'ematiques, Universit\'e Paris 7 \\ {\it 2 Place Jussieu 75251 Paris
 cedex 05, France.}\\E Mail: finkel@logique.jussieu.fr }
\date{}
\begin{document}

\newtheorem{The}{Theorem}[section]
\newtheorem{Pro}[The]{Proposition}
\newtheorem{Deff}[The]{Definition}
\newtheorem{Lem}[The]{Lemma}
\newtheorem{Rem}[The]{Remark}
\newtheorem{Exa}[The]{Example}
\newtheorem{Cor}[The]{Corollary}

\newcommand{\vp}{\varphi}
\newcommand{\lb}{\linebreak}
\newcommand{\fa}{\forall}
\newcommand{\Ga}{\Gamma}
\newcommand{\Gas}{\Gamma^\star}
\newcommand{\Gao}{\Gamma^\omega}
\newcommand{\Si}{\Sigma}
\newcommand{\Sis}{\Sigma^\star}
\newcommand{\Sio}{\Sigma^\omega}
\newcommand{\ra}{\rightarrow}
\newcommand{\hs}{\hspace{12mm}

\noi}
\newcommand{\lra}{\leftrightarrow}
\newcommand{\la}{language}
\newcommand{\ite}{\item}
\newcommand{\Lp}{L(\varphi)}
\newcommand{\abs}{\{a, b\}^\star}
\newcommand{\abcs}{\{a, b, c \}^\star}
\newcommand{\ol}{ $\omega$-language}
\newcommand{\orl}{ $\omega$-regular language}
\newcommand{\om}{\omega}
\newcommand{\nl}{\newline}
\newcommand{\noi}{\noindent}
\newcommand{\tla}{\twoheadleftarrow}
\newcommand{\de}{deterministic }
\newcommand{\proo}{\noi {\bf Proof.} }
\newcommand {\ep}{\hfill $\square$}

\maketitle

\begin{abstract}
\noi $\omega$-powers of finitary languages  
are \ol s in the form $V^\om$, where $V$ is a finitary language over 
a finite  alphabet $\Si$.  
 Since the set $\Sio$ of infinite words over $\Si$ can be  equipped 
with the usual Cantor topology, the question of the topological complexity 
of $\om$-powers  naturally arises and  has been raised by 
Niwinski \cite{Niwinski90}, by Simonnet \cite{Simonnet92},  and by Staiger \cite{Staiger97b}. 
It has been  proved  in \cite{Fin01a} that 
 for each integer $n\geq 1$, there exist some 
$\om$-powers of  context free languages 
which are ${\bf \Pi_n^0}$-complete Borel sets, and in \cite{Fin03a} that there exists a 
context free language $L$ such that $L^\om$ is analytic but not Borel. 
But the question was still open whether there exists a finitary language $V$ 
such that $V^\om$ is a Borel set of infinite rank. 
\nl We answer  this question in this paper, giving an example of  a finitary language 
whose $\om$-power is Borel of infinite rank.   
\end{abstract}

\noi {\small {\bf Keywords:} Infinite words; 
$\omega$-languages; $\omega$-powers; Cantor topology; 
topological complexity; Borel sets; infinite rank.}

\section{Introduction}

\hs $\omega$-powers are \ol s in the form $V^\om$, where $V$ is a finitary language. 
The operation $V \ra V^\om$  is a fundamental operation over finitary languages 
leading to \ol s. This operation  appears  in the characterization of the class 
$REG_\om$  of  \orl s (respectively,  of the class $CF_\om$ of context free \ol s) 
 as the $\om$-Kleene closure 
of the family $REG$ of regular finitary languages (respectively,   of the 
family $CF$ of context free finitary languages) \cite{Staiger97}.  
\nl The set $\Sio$ of infinite words over a finite alphabet $\Si$ is usually equipped 
with the Cantor topology which may be defined by a distance, see \cite{Staiger97}
 \cite{PerrinPin}. 
One can then study the complexity of \ol s, i.e.  languages of infinite words, 
by considering their topological complexity, 
with regard to the Borel hierarchy (and beyond to the projective hierarchy) \cite{Staiger97} 
\cite{PerrinPin}. 
\nl  The question of the topological complexity of $\om$-powers of 
finitary  languages naturally arises.  It has been posed by 
Niwinski \cite{Niwinski90}, by Simonnet \cite{Simonnet92} and by Staiger \cite{Staiger97b}. 
The $\om$-power of a finitary language is always an analytic set because 
it is the continuous image of a compact set $\{0,1, \ldots ,n\}^\om$ for $n\geq 0$
or of the Baire space $\om^\om$, 
\cite{Simonnet92} \cite{Fin01a}. 
It has been  proved  in \cite{Fin01a} that 
for each integer $n\geq 1$, there exist some $\om$-powers of  context free languages 
which are ${\bf \Pi_n^0}$-complete Borel sets, and in \cite{Fin03a} that there exists a 
context free language $L$ such that $L^\om$ is analytic but not Borel. 
\nl  But the question was still open whether there exists a finitary language $V$ 
such that $V^\om$ is a Borel set of infinite rank. 
\nl We answer  this question in this paper, giving an example of  a finitary language 
whose $\om$-power is Borel of infinite rank.   
\nl  The paper is organized as follows. In Section 2 we recall definitions of Borel sets 
and previous results and we proved our main result in Section 3. 

\section{Recall on Borel sets and previous results} 

We assume the reader to be familiar with the theory of formal \ol s  
\cite{Thomas90}, \cite{Staiger97}.
We shall use usual notations of formal language theory. 

\hs  When $\Si$ is a finite alphabet, a {\it non-empty finite word} over $\Si$ is any 
sequence $x=a_1\ldots a_k$ , where $a_i\in\Sigma$ 
for $i=1,\ldots ,k$ , and  $k$ is an integer $\geq 1$. The {\it length}
 of $x$ is $k$, denoted by $|x|$.
 The {\it empty word} has no letter and is denoted by $\lambda$; its length is $0$. 
 For $x=a_1\ldots a_k$, we write $x(i)=a_i$  
and $x[i]=x(1)\ldots x(i)$ for $i\leq k$ and $x[0]=\lambda$.
 $\Sis$  is the {\it set of finite words} (including the empty word) over $\Sigma$.
 
 \hs  The {\it first infinite ordinal} is $\om$.
 An $\om$-{\it word} over $\Si$ is an $\om$ -sequence $a_1 \ldots a_n \ldots$, where for all 
integers $ i\geq 1$, ~
$a_i \in\Sigma$.  When $\sigma$ is an $\om$-word over $\Si$, we write
 $\sigma =\sigma(1)\sigma(2)\ldots \sigma(n) \ldots $,  where for all $i$,~ $\sigma(i)\in \Si$,
and $\sigma[n]=\sigma(1)\sigma(2)\ldots \sigma(n)$  for all $n\geq 1$ and $\sigma[0]=\lambda$.

 \hs  The {\it prefix relation} is denoted $\sqsubseteq$: a finite word $u$ is a {\it prefix} 
of a finite word $v$ (respectively,  an infinite word $v$), denoted $u\sqsubseteq v$,  
 if and only if there exists a finite word $w$ 
(respectively,  an infinite word $w$), such that $v=u.w$.
 The {\it set of } $\om$-{\it words} over  the alphabet $\Si$ is denoted by $\Si^\om$.
An  $\om$-{\it language} over an alphabet $\Sigma$ is a subset of  $\Si^\om$.
 
\hs  For $V\subseteq \Sis$, the $\om$-{\it power} of $V$ is the \ol:
$$V^\om = \{ \sigma =u_1\ldots u_n \ldots \in \Si^\om \mid 
 \fa i\geq 1 ~ u_i\in V-\{\lambda\} \}$$

\noi We assume the reader to be familiar with basic notions of topology which
may be found in \cite{Moschovakis80} \cite{LescowThomas} \cite{Kechris94} 
\cite{Staiger97} \cite{PerrinPin}.
For a finite alphabet $X$, we consider $X^\om$ 
as a topological space with the Cantor topology.
The {\it open sets} of $X^\om$ are the sets of the form $W.X^\om$, where $W\subseteq X^\star$.
A set $L\subseteq X^\om$ is a {\it closed set} iff its complement $X^\om - L$ is an open set.
Define now the {\it Borel Hierarchy} of subsets of $X^\om$:

\begin{Deff}
For a non-null countable ordinal $\alpha$, the classes ${\bf \Si^0_\alpha }$
 and ${\bf \Pi^0_\alpha }$ of the Borel Hierarchy on the topological space $X^\om$ 
are defined as follows:
\nl ${\bf \Si^0_1 }$ is the class of open subsets of $X^\om$.
\nl ${\bf \Pi^0_1 }$ is the class of closed subsets of $X^\om$.
\nl and for any countable ordinal $\alpha \geq 2$: 
\nl ${\bf \Si^0_\alpha }$ is the class of countable unions of subsets of $X^\om$ in 
$\cup_{\gamma <\alpha}{\bf \Pi^0_\gamma }$.
 \nl ${\bf \Pi^0_\alpha }$ is the class of countable intersections of subsets of $X^\om$ in 
$\cup_{\gamma <\alpha}{\bf \Si^0_\gamma }$.
\end{Deff}

\noi For 
a countable ordinal $\alpha$,  a subset of $X^\om$ is a Borel set of {\it rank} $\alpha$ iff 
it is in ${\bf \Si^0_{\alpha}}\cup {\bf \Pi^0_{\alpha}}$ but not in 
$\bigcup_{\gamma <\alpha}({\bf \Si^0_\gamma }\cup {\bf \Pi^0_\gamma})$.

\hs There are also some subsets of $X^\om$ which are not Borel.  In particular 
the class of Borel subsets of $X^\om$ is strictly included into 
the class  ${\bf \Si^1_1}$ of {\it analytic sets} which are 
obtained by projection of Borel sets, 
see for example \cite{Staiger97} \cite{LescowThomas}  \cite{PerrinPin} \cite{Kechris94}
 for more details.

\hs  We now define completeness with regard to reduction by continuous functions. 
For a countable ordinal  $\alpha\geq 1$, a set $F\subseteq X^\om$ is said to be 
a ${\bf \Si^0_\alpha}$  
(respectively,  ${\bf \Pi^0_\alpha}$, ${\bf \Si^1_1}$)-{\it complete set} 
iff for any set $E\subseteq Y^\om$  (with $Y$ a finite alphabet): 
 $E\in {\bf \Si^0_\alpha}$ (respectively,  $E\in {\bf \Pi^0_\alpha}$,  $E\in {\bf \Si^1_1}$) 
iff there exists a continuous function $f: Y^\om \ra X^\om$ such that $E = f^{-1}(F)$. 
 ${\bf \Si^0_n}$
 (respectively ${\bf \Pi^0_n}$)-complete sets, with $n$ an integer $\geq 1$, 
 are thoroughly characterized in \cite{Staiger86a}.  

\hs In particular  $\mathcal{R}=(0^\star.1)^\om$  
is a well known example of 
${\bf \Pi^0_2 }$-complete subset of $\{0, 1\}^\om$. It is the set of 
$\om$-words over $\{0, 1\}$ having infinitely many occurrences of the letter $1$. 
Its  complement 
$\{0, 1\}^\om - (0^\star.1)^\om$ is a 
${\bf \Si^0_2 }$-complete subset of $\{0, 1\}^\om$.

\hs We shall recall the  definition of the operation $A \ra A^\approx$ over sets of 
 infinite words we introduced in \cite{Fin01a} and which is a 
simple variant of Duparc's operation of exponentiation 
$A \ra A^\sim$ \cite{Duparc01}. 
\nl For a finite alphabet $\Si$ 
we denote $\Si^{\leq \om}=\Sio \cup \Sis$. Let now $\tla$ a letter not in $\Si$ and  
$X=\Si\cup \{\tla\}$. 
For $x \in X^{\leq \om}$, $x^\tla$ denotes the string $x$, once every $\tla$ occuring in $x$
has been ``evaluated" to the back space operation ( the one familiar to your computer!),
proceeding from left to right inside $x$. In other words $x^\tla = x$ from which every
 interval of the form $`` a\tla "$ ($a\in \Si$) is removed. We add the convention that 
$(u.\tla)^\tla$ is undefined if $|u^\tla|=0$, i.e. when the last letter $\tla$ can not be used 
as an eraser (because every letter of $\Si$ in $u$ 
has already been erased by some erasers $\tla$ placed  in $u$). 
Remark that the resulting word $x^\tla$ may be 
finite or infinite.

\hs For example if  $u=(a\tla)^n$, for $n\geq 1$, or 
$u=(a\tla)^\om$ then $(u)^\tla=\lambda$,
\nl if $u=(ab\tla)^\om$ then $(u)^\tla=a^\om$,
 if $u=bb(\tla a)^\om$ then $(u)^\tla=b$,  
\nl if $u=\tla(a\tla)^\om$ or $u=a\tla\tla a^\om$ or $u=(a\tla\tla)^\om$ 
then $(u)^\tla$ is undefined.

\begin{Deff}
For $A\subseteq \Si^{\om}$, ~~~~
 $A^\approx =\{x\in (\Si\cup \{\tla\})^{\om} \mid  x^\tla\in A\}. $
\end{Deff}

\noi  The following result follows easily from \cite{Duparc01} and  was applied  
in \cite{Fin01a} to study the $\om$-powers of finitary context free languages.

\begin{The}\label{thedup}
Let $n$ be an  integer $\geq 2$ and  $A\subseteq \Si^\om$ be a ${\bf \Pi_n^0}$-complete
 set. Then $A^\approx$ is a 
${\bf \Pi_{n+1}^0}$-complete subset of $(\Si \cup\{\tla\})^\om$.
\end{The}

\noi For each \ol~ $A \subseteq \Si^{\om}$, the 
\ol~ $A^\approx$ can be easily described from $A$ by the use of the notion of substitution 
which we recall now. 
\nl A {\it substitution}  is defined by a mapping 
$f: \Si\ra P(\Ga^\star)$, where $\Si =\{a_1, \ldots ,a_n\}$  and $\Ga$ are two finite alphabets, 
$f: a_i \ra L_i$ where for all integers $i\in [1;n]$, $f(a_i)=L_i$ is a finitary language 
over the alphabet  $\Ga$.
\nl Now this mapping is extended in the usual manner to finite words:
$f(a_{i_1} \ldots a_{i_n})= L_{i_1} \ldots L_{i_n}$,   
and to finitary languages $L\subseteq \Sis$: 
$f(L)=\cup_{x\in L} f(x)$.  
\noi If for each integer $i\in [1;n]$ the language  $L_i$ does not 
contain the empty word, then the mapping $f$ may be extended to $\om$-words:
$$f(x(1)\ldots x(n)\ldots )= \{u_1\ldots u_n \ldots  \mid \fa i \geq 1 \quad u_i\in f(x(i))\}$$ 
\noi and to \ol s $L\subseteq \Sio$ by setting  $f(L)=\cup_{x\in L} f(x)$.   

\hs  Let $L_1 = \{ w\in  (\Si \cup\{\tla\})^\star \mid w^\tla =\lambda \}$.  $L_1$ is a 
context free (finitary) 
language generated by the context free grammar with the following production rules:
\nl $S\ra aS\tla S$  with $a\in \Si$; 
and  $S\ra \lambda$ (where $ \lambda$ is the empty word). 
\nl 
Then, for each $\om$-language $A \subseteq \Si^{\om}$, the  $\om$-language 
$A^\approx \subseteq (\Si\cup\{\tla\})^\om$
is obtained by substituting in $A$ the language $L_1.a$ for each letter $a\in \Si$. 

\hs By definition  the  operation $A \ra A^\approx$ conserves the $\om$-powers 
of finitary languages. Indeed if $A=V^\om$ for some language $V   \subseteq \Sis$ then 
$A^\approx=g(V^\om)=( g(V) )^\om$ where $g: \Si \ra P((\Si \cup\{\tla\})^\star)$ is the 
substitution defined by 
$g(a)=L_1.a$ for every letter $a\in \Si$.

\section{An $\om$-power which is Borel of infinite rank}

\noi We can now iterate $k$ times this operation $A \ra A^\approx$. 
 More precisely, we define, for a set $A\subseteq \Si^{\om}$:
\nl $A_k^{\approx .0}=A$, 
\nl $A_k^{\approx .1}=A^\approx$, 
\nl $A_k^{\approx .2}=(A_k^{\approx .1})^\approx$, 
and
\nl $A_k^{\approx .k}=(A_k^{\approx .(k-1)})^\approx$, 
\nl  where we apply $k$ times the operation $A\ra A^\approx$ 
with different new letters $\tla_{k}$, $\tla_{k-1}$, \ldots ,$\tla_3$, $\tla_2$,  $\tla_1$,   
in such a way that we have 
successively: 

\hs $A_k^{\approx .0}=A\subseteq \Si^{\om}$, 
\nl $A_k^{\approx .1}\subseteq (\Si\cup\{\tla_k\})^{\om}$,   
\nl $A_k^{\approx .2} \subseteq (\Si\cup\{\tla_k, \tla_{k-1}\})^{\om}$,  
\nl   \ldots \ldots \ldots \ldots
\nl $A_k^{\approx .k} \subseteq 
(\Si \cup\{\tla_k, \tla_{k-1}, \ldots , \tla_1\})^{\om}$. 

\hs  For a reason which will be clear later we have chosen to  successively call the erasers 
$\tla_{k}$, $\tla_{k-1}$, \ldots ,   $\tla_2$, $\tla_1$, in this precise order. 
 We set now $A^{\approx .k} = A_k^{\approx .k}$ so it holds that 
 $$A^{\approx .k} \subseteq 
(\Si \cup\{\tla_k, \tla_{k-1}, \ldots , \tla_1\})^{\om}$$
\noi Notice that definitions of  $A_k^{\approx .1}, 
 A_k^{\approx .2},  \ldots , A_k^{\approx .(k-1)}$  were just some 
intermediate steps for the definition of  $A^{\approx .k}$     and will not be used later. 

\hs We can also describe the operation $A \ra A^{\approx .k}$ in a similar manner as 
in the case  of the operation $A \ra A^{\approx}$, by the use of the notion of substitution. 
 
\hs  Let $L_k \subseteq (\Si\cup \{\tla_k, \tla_{k-1}, \ldots , \tla_1\})^\star$
be the  language containing  (finite) words $u$, such that all letters of $u$ 
have been erased when the operations of erasing using the erasers $\tla_1$,  $\tla_2$, \ldots ,
$\tla_{k-1}$, $\tla_{k}$,  are successively 
applied to $u$. 
\nl Notice that  the operations 
of erasing have to be  done in a good order:  the first operation of erasing uses the 
eraser $\tla_1$, then the second one uses the eraser $\tla_{2}$, and so on \ldots 
\nl So an eraser $\tla_j$ may only erase a letter of $\Si$ or an other ``eraser" $\tla_i$ for 
some integer $i>j$. 
\nl It is  easy to see that $L_k$ is a context free language. In fact $L_k$ belongs to the 
subclass of iterated counter languages which is the closure under substitution 
of the class of one counter languages, see \cite{ABB96} \cite{Fin01a} for more details.

\hs 
Let now $h_k$  be the substitution: 
$\Si \ra P((\Si \cup\{\tla_k, \tla_{k-1}, \ldots , \tla_1\})^\star)$ defined by 
 $h_k(a)=L_k.a$ for every letter $a\in \Si$. 
\nl 
Then it holds that, for $A \subseteq \Sio$, 
 $A^{\approx .k} = h_k(A)$,  i.e. $A^{\approx .k}$
is obtained by substituting in $A$ the language $L_k.a$ for each letter $a\in \Si$.

\hs The \ol~  $\mathcal{R} = (0^\star.1)^\om = \mathcal{V}^\om$, where 
$\mathcal{V} = (0^\star.1)$, 
 is  ${\bf \Pi_2^0}$-complete.  Then, 
by Theorem \ref{thedup},   
 for each integer $p \geq 1$,  
$h_{p}(\mathcal{V}^\om)=(h_{p}(\mathcal{V}))^\om$ 
 is a ${\bf \Pi_{p+2}^0}$-complete set. 

\hs We can see that the languages $L_k$, for $k\geq 1$, form a sequence which is strictly 
increasing for the inclusion relation: 
$$L_1 \subset L_2 \subset L_3 \subset \ldots  \subset L_i \subset L_{i+1} \ldots$$
\noi In order to construct some $\om$-power which is Borel of infinite rank, we could try 
to substitute the language $\cup_{k\geq 1} L_k.a$ to each letter $a\in \Si$. 
 But the language $\cup_{k\geq 1} L_k.a$ is defined over the infinite alphabet 
 $\Si \cup\{\tla_1, \tla_2, \tla_3, \ldots\}$, so we shall first 
 code every eraser $\tla_j$ by a finite word over a fixed finite alphabet. 
The eraser $\tla_j$ will be coded by the finite word 
$\alpha.\beta^j.\alpha$ over the alphabet $\{\alpha, \beta\}$, where $\alpha$ and $\beta$ are 
two new letters.  

\hs  The morphism 
$\vp_p:~  (\Si\cup \{\tla_1, \ldots , \tla_p\})^\star  \ra  
(\Si \cup \{\alpha, \beta\})^\star$
defined  by  $\vp_p(c)=c$ for each $c\in \Si$ and $\vp_p(\tla_j)=\alpha.\beta^j.\alpha$ 
for each integer 
$j \in [1, p]$, can be  naturally extended  to a continuous function 
$\psi_p:~  (\Si\cup \{\tla_1, \ldots , \tla_p\})^\om  \ra
 (\Si \cup \{\alpha, \beta\})^\om$. 
 
\hs Let now 
$$\mathcal{L}=\bigcup_{n\geq 1}\vp_n(L_n)$$
\noi and $h: \Si \ra P((\Si \cup \{\alpha, \beta\})^\star)$ 
 be the substitution defined by 
$$h(a)=\mathcal{L}.a$$
\noi for each $a\in \Si$. 

\hs We can now state our main result: 

\begin{The}\label{the}
Let $\mathcal{V} = (0^\star.1)$. Then the $\om$-power $( h(\mathcal{V}) )^\om 
 \subseteq \{0, 1, \alpha, \beta\}^\om$ 
is a Borel set of infinite rank. 
\end{The}

\noi To prove this theorem, we  shall proceed by successive lemmas. 

\begin{Lem}\label{lemm}
For all integers $p\geq 1$, the  
\ol~  $\psi_p(\mathcal{R}^{\approx .p})$  is a ${\bf \Pi_{p+2}^0}$-complete subset of 
$(\Si \cup \{\alpha, \beta\})^\om$.
\end{Lem}

\proo  First we prove that $\psi_p(\mathcal{R}^{\approx .p})$  
is in the class ${\bf \Pi_{p+2}^0}$. 
\nl For $\Si=\{0, 1\}$, 
$\psi_p((\Si\cup \{\tla_1, \ldots , \tla_p\})^\om )$ is the continuous image by 
$\psi_p$ of the compact set $(\Si\cup \{\tla_1, \ldots , \tla_p\})^\om$, hence it is also 
a  compact set. 
\nl The function $\psi_p$ is injective and continuous thus it induces an 
homeomorphism $\psi'_p$ between the two compact sets 
$(\Si\cup \{\tla_1, \ldots , \tla_p\})^\om $ 
and $\psi_p((\Si\cup \{\tla_1, \ldots , \tla_p\})^\om )$. 
\nl   We have already seen that, for each integer  $p\geq 1$, the  
\ol~  $\mathcal{R}^{\approx .p}$ is a ${\bf \Pi_{p+2}^0}$-complete subset of 
$(\Si\cup \{\tla_1, \ldots , \tla_p\})^\om $. 
  Then $\psi'_p(\mathcal{R}^{\approx .p})$ 
is a ${\bf \Pi_{p+2}^0}$-subset of $\psi_p((\Si\cup \{\tla_1, \ldots , \tla_p\})^\om )$, 
because the function $\psi'_p$ is an  homeomorphism. 
\nl  But one can prove, 
by induction over the integer j$\geq 1$, that  each 
${\bf \Pi_{j}^0}$ 
subset $K$ 
of $\psi_p((\Si\cup \{\tla_1, \ldots , \tla_p\})^\om )$ is also a ${\bf \Pi_{j}^0}$ 
subset of 
$(\Si \cup \{\alpha, \beta\})^\om$.  
Thus 
$\psi'_p(\mathcal{R}^{\approx .p})=\psi_p(\mathcal{R}^{\approx .p})$ 
is a ${\bf \Pi_{p+2}^0}$-subset of  
$(\Si \cup \{\alpha, \beta\})^\om$. 
\nl  Remark  now  that 
 the set $\mathcal{R}^{\approx .p}$ being ${\bf \Pi_{p+2}^0}$-complete, every 
${\bf \Pi_{p+2}^0}$-subset of $X^\om$,  for $X$ a finite alphabet, is the inverse image 
of $\mathcal{R}^{\approx .p}$ by a continuous function. But it holds that 
$\mathcal{R}^{\approx .p} = \psi_p^{-1}(\psi_p(\mathcal{R}^{\approx .p}))$, 
where $\psi_p$ is a continuous function.  Thus every 
${\bf \Pi_{p+2}^0}$-subset of $X^\om$,   for $X$ a finite alphabet, is 
the inverse image 
of $\psi_p(\mathcal{R}^{\approx .p})$ by a continuous function. Therefore  
$\psi_p(\mathcal{R}^{\approx .p})$ is also a ${\bf \Pi_{p+2}^0}$-complete subset of 
$(\Si \cup \{\alpha, \beta\})^\om$.  
\ep

\begin{Lem}\label{not-fin}
The set $( h(\mathcal{V}) )^\om$ is not a Borel set 
of finite rank.  
\end{Lem} 

\proo Consider, for each integer $p\geq 1$, the regular \ol~ 
$$R_p = \psi_p(\{0, 1, \tla_1, \tla_2, \ldots , \tla_p\}^\om ) = 
\{0, 1, \alpha.\beta.\alpha, \alpha.\beta^2.\alpha, \ldots , \alpha.\beta^p.\alpha\}^\om$$ 
\noi We have seen that $R_p$ is compact hence it is a closed set. 
And by construction it holds that 
$( h(\mathcal{V}) )^\om \cap R_p = \psi_p (  h_p ( \mathcal{V}^\om ) ) = 
\psi_p (  \mathcal{R}^{\approx.p} )$
 where $\mathcal{R}=\mathcal{V}^\om$, so by Lemma \ref{lemm} 
this set is a ${\bf \Pi_{p+2}^0}$-complete subset of 
$\{0, 1, \alpha, \beta\}^\om$. 

\hs If $( h(\mathcal{V}) )^\om$ was a Borel set of finite rank it would be in the class 
${\bf \Pi_{J}^0}$ for some integer $J\geq 1$. But then $( h(\mathcal{V}) )^\om \cap R_p$ 
would be the intersection of a ${\bf \Pi_{J}^0}$-set and of a closed, 
i.e. ${\bf \Pi_{1}^0}$-set. Thus, for each integer $p\geq 1$, the set 
 $( h(\mathcal{V}) )^\om \cap R_p$ would be a ${\bf \Pi_{J}^0}$-set. This would lead to a
contradiction because,  for  $J=p$, a ${\bf \Pi_{J}^0}$-set cannot be a 
${\bf \Pi_{p+2}^0}$-complete set.   
\ep 

\begin{Lem}\label{code}
Every $\om$-word 
$x \in ( h(\mathcal{V}) )^\om$ has a unique 
decomposition of the form $x = u_1.u_2 \ldots u_n \ldots $ where 
for all $i\geq 1$~ $u_i \in h(\mathcal{V})$. 
\end{Lem} 
 
\proo Towards a contradiction assume on the contrary that some 
$\om$-word  $x\in ( h(\mathcal{V}) )^\om = ( h(0^\star.1) )^\om$ has (at least) two distinct 
decompositions  in words of $h(\mathcal{V})$. 
So there are some words $u_j, u'_j \in h(\mathcal{V})$, for $j\geq 1$, such that  
$$x = u_1.u_2 \ldots u_n \ldots = u'_1.u'_2 \ldots u'_n \ldots $$ 
\noi and an integer $J\geq 1$ such that $u_j=u'_j$ for $j<J$ and $u_J  \sqsubset u'_J$, i.e. 
$u_J$ is a strict prefix of $u'_J$.
 Then for each 
integer $j\geq 1$, there are integers $n_j, n'_j \geq 0$ such that $u_j \in h(0^{n_j}.1)$ and 
$u'_j \in h(0^{n'_j}.1)$. Thus there
are some finite words $v_i^j \in \mathcal{L}$, where $i$ is 
an integer in $[1, n_j+1]$, and  $w_i^j  \in \mathcal{L}$,  where $i$ is 
an integer in $[1, n'_j+1]$, such 
that 
$$u_j = v_1^j.0.v_2^j.0 \ldots v_{n_j}^j.0.v_{n_j+1}^j.1 ~~\mbox{  and } ~~ 
u'_j = w_1^j.0.w_2^j.0 \ldots w_{n'_j}^j.0.w_{n'_j+1}^j.1 $$
\noi We consider now $x$ given by its first decomposition $x = u_1.u_2 \ldots u_n \ldots$ 
\nl Let  now  $x^{(1)}$ be the $\om$-word obtained from $x$ by using the (code of the) eraser 
$\tla_1$ as an eraser which may erase letters $0$, $1$, and (codes of the) erasers  $\tla_p$ 
for $p>1$.
 Remark that by construction  these operations of erasing occur inside the words 
$v_i^j$ for $j\geq 1$ and  $i \in [1, n_j+1]$. 
\nl Next let $x^{(2)}$ be the $\om$-word obtained from $x^{(1)}$ 
 by using the (code of the) eraser 
$\tla_2$ as an eraser which may erase letters $0$, $1$, and (codes of the) erasers  $\tla_p$ 
for $p>2$.
Again these erasing operations occur 
inside the words $v_i^j$. 
\nl We can now iterate this process. 
 Assume that, after having successively used the erasers 
$\tla_1$, $\tla_2$, \ldots , $\tla_n$, 
for some integer $n\geq 1$, we have got the $\om$-word $x^{(n)}$ from the $\om$-word $x$. 
 We can now define $x^{(n+1)}$ as the $\om$-word obtained from $x^{(n)}$ by using the 
(code of the) eraser $\tla_{n+1}$ as an eraser which may erase 
letters $0$, $1$, and (codes of the) erasers  $\tla_p$ for $p>n+1$.

\hs We  shall denote $K_{(i, j)}=min \{k \geq 1 \mid v_i^j \in \vp_k(L_k) \}$. Then  
 $v_i^j \in \vp_{K_{(i, j)}}(L_{K_{(i, j)}})$ for all integers
 $j\geq 1$ and  $i \in [1, n_j+1]$. 
Thus after $K_j=max \{ K_{(i, j)} \mid i \in [1, n_j+1] \}$ 
steps all words $v_i^j$, for $i \in [1, n_j+1]$, 
have been completely erased and, from the finite word 
$u_j$, it remains only the finite word $0^{n_j}.1$.  
\nl After $K^J=max\{ K_j \mid j \in [1, J] \}$ steps, from the word $u_1.u_2 \ldots u_J$, it 
remains the word $0^{n_1}.1.0^{n_2}.1.0^{n_3}.1 \ldots 0^{n_J}.1$ which is a strict prefix of 
$x^{(K^J)}$. In particular the $J$-th letter $1$ of $x^{(K^J)}$ is the last letter of $u_J$, 
which has not been erased,  and 
it will not be erased by next erasing operations. 

\hs  Notice that the successive erasing operations are in fact 
applied to  $x$ independently 
of the decomposition of $x$ in words of  $h(\mathcal{V})$. 
So consider now the above  erasing operations applied to   $x$ given by its second decomposition. 
\nl Let 
$K'_{(i, j)}=min \{k \geq 1 \mid w_i^j \in \vp_k(L_k) \}$, 
and 
$K^{'J}=max\{  K'_{(i, j)} \mid  j \in [1, J] \mbox{ and } i \in [1, n'_j+1]  \}$. 
$K^j=K^{'j}$  holds  for $1\leq j <J$ and  $K^J \leq K^{'J}$. 
\nl We see that, 
after $K^{'J}$ steps, 
the word $0^{n_1}.1.0^{n_2}.1.0^{n_3}.1 \ldots 0^{n_{J-1}}.1.0^{n'_J}.1$  is a strict prefix of 
$x^{(K^{'J})}$. The $J$-th letter $1$ of $x^{(K^{'J})}$ is the last letter of $u'_J$ 
(which has not  been erased). But we have seen above that it is also the last letter of 
$u_J$ (which has not  been erased). 
\nl Thus we would have $u_J=u'_J$ and this leads to a contradiction. 
\ep 

\hs Remark that in terms  of code theory Lemma \ref{code} states  that 
the language  $h(\mathcal{V})$ is an $\om$-code.

\begin{Lem}\label{borel} 
The  set $( h(\mathcal{V}) )^\om$ is  a Borel set.
\end{Lem} 

\proo Assume on the contrary that $( h(\mathcal{V}) )^\om$ is  an analytic but non Borel set. 
Recall that lemma 4.1 of \cite{Fink-Sim} states that if 
$X$ and $Y$ are  finite alphabets having at least two letters and 
 $B$ is a Borel subset of 
$X^\om \times Y^\om$ such that 
$PROJ_{X^\om}(B)=\{\sigma \in X^\om \mid  \exists \nu ~(\sigma, \nu) \in B\}$ 
is not  Borel, 
then there are $2^{\aleph_0}$ $\om$-words  $\sigma \in X^\om$ such that 
$B_\sigma=\{\nu \in Y^\om \mid (\sigma,\nu) \in B\}$ 
has cardinality $2^{\aleph_0}$ 
(where $2^{\aleph_0}$ is 
the cardinal of the continuum).
\nl We can now reason as in the proof of 
Fact 4.5 in \cite{Fink-Sim}. 
\nl Let $\theta$ be a  recursive  enumeration  of the set $h(\mathcal{V})$. 
The  function $\theta : \mathbb{N} \ra h(\mathcal{V})$ 
is a bijection  and we denote $u_i=\theta(i)$. 
\nl Let now $\mathcal{D}$ be the set of pairs 
$(\sigma, \nu) \in \{0, 1\}^\om \times \{0, 1, \alpha, \beta \}^\om$ such that:  

\begin{enumerate} 
\ite $\sigma \in (0^\star.1)^\om$, so $\sigma$ may be written in the form 
$$\sigma = 0^{n_1}.1.0^{n_2}.1.0^{n_3}.1 \ldots 0^{n_p}.1.0^{n_{p+1}}.1 \ldots $$
\noi where $\fa i \geq 1$ ~ $n_i \geq 0$, ~and  
\ite $$\nu = u_{n_1}.u_{n_2}.u_{n_3}\ldots u_{n_p}.u_{n_{p+1}} \ldots $$
\end{enumerate} 

\noi $\mathcal{D}$ is a Borel subset of  $\{0, 1\}^\om \times \{0, 1, \alpha, \beta \}^\om$ 
because it is accepted by a deterministic Turing machine with a B\"uchi acceptance condition 
\cite{Staiger97}. 
\nl But $PROJ_{\{0, 1, \alpha, \beta \}^\om}(\mathcal{D})= (h(\mathcal{V}))^\om$ 
would be a non Borel set thus there would be  $2^{\aleph_0}$ $\om$-words $\nu$ in 
$(h(\mathcal{V}))^\om$    such that $\mathcal{D}_\nu$ has cardinality $2^{\aleph_0}$. 
This means that there would  exist $2^{\aleph_0}$ $\om$-words  
$\nu \in  (h(\mathcal{V}))^\om$ having 
 $2^{\aleph_0}$  decompositions  in words in $h(\mathcal{V})$. 
\nl This would lead to a (strong)  contradiction with Lemma \ref{code}. 
\ep 

\hs Theorem \ref{the} follows now directly from Lemmas \ref{not-fin} and \ref{borel}. 

\section{Concluding Remarks}

\noi  We already knew that there are $\om$-powers of every finite Borel rank \cite{Fin01a}. 
We have proved that there exists some $\om$-powers of infinite Borel rank. 
The language  $h(\mathcal{V})$ is very simple to describe. It is obtained by substituting in 
the regular language  $\mathcal{V} = (0^\star.1)$ the language $\mathcal{L}.a$ to each 
letter $a\in \{0, 1\}$, where $\mathcal{L}=\bigcup_{n\geq 1}\vp_n(L_n)$.  Notice that the 
language $\mathcal{L}$ is not context free but it is the union of the increasing sequence of 
context free languages $\vp_n(L_n)$. Then $\mathcal{L}$ is a very simple recursive language 
and so is $h(\mathcal{V})$. 
\nl The question is left open whether there is a context free language $W$ such that 
$W^\om$ is Borel of infinite rank. 
\nl  The question also  naturally arises 
to know what are all the possible infinite Borel ranks of 
$\om$-powers of finitary languages or of finitary  
languages belonging to some natural class like the 
class of context free languages (respectively,   languages 
accepted by stack automata,  recursive languages, 
 recursively enumerable languages,  \ldots ). 

\hs {\bf  Acknowledgements.}  We thank the anonymous referee 
for useful comments on a preliminary version of this paper.

\end{document}